\documentclass[notitlepage,nofootinbib,preprintnumbers,eqsecnum,amssymb]{revtex4-1}

\usepackage{amsfonts,amssymb,amsmath,graphicx,color,bm}
\definecolor{ultramarine}{rgb}{0.07, 0.04, 0.56}
\definecolor{cadmiumgreen}{rgb}{0.0, 0.42, 0.24}
\definecolor{indigo(dye)}{rgb}{0.0, 0.25, 0.42}
\usepackage[linktocpage=true]{hyperref}
\hypersetup{
colorlinks=true,
linkcolor=cadmiumgreen,
citecolor=ultramarine,
urlcolor=indigo(dye),
}
\usepackage{caption}
\usepackage{subcaption}
\usepackage{slashed} 
\usepackage{enumerate} 

\newcommand{\f}[2]{\frac{#1}{#2}}
\newcommand{\mk}[1]{\left( #1 \right)}
\newcommand{\kk}[1]{\left[ #1 \right]}
\newcommand{\ck}[1]{\left\{ #1 \right\}}
\newcommand{\be}{\begin{equation}}
\newcommand{\ee}{\end{equation}}
\def\Sp{\epsilon_s}
\def\tSp{\tilde{\epsilon}_s}
\def\R{\mathcal{R}_c}
\def\tR{\tilde{\mathcal R}_c}
\def\kA{{({\rm A})}}
\def\kB{{({\rm B})}}
\def\kH{{({\rm H})}}
\def\kHA{{({\rm H}\to{\rm A})}}
\def\kHB{{({\rm H}\to{\rm B})}}
\def\kAB{{({\rm A}\to{\rm B})}}
\def\gmn{g_{\mu\nu}}
\def\pmn{\partial_\mu \phi \partial_\nu \phi}

\begin{document}

\title{Disformal invariance of curvature perturbation}

\author{Hayato Motohashi}
\email{motohashi(at)kicp.uchicago.edu}
\affiliation{Kavli Institute for Cosmological Physics, The University of Chicago, Chicago, Illinois 60637, U.S.A.}
\author{Jonathan White}
\email{jwhite(at)post.kek.jp}
\affiliation{Research Center for the Early Universe (RESCEU),
Graduate School of Science, The University of Tokyo, Tokyo 113-0033, Japan}
\affiliation{Theory Center, KEK, Tsukuba 305-0801, Japan}
\preprint{RESCEU-8/15,~KEK-TH-1802,~KEK-Cosmo-167}

\begin{abstract}
We show that under a general disformal transformation the linear
comoving curvature perturbation is not identically invariant, but is
invariant on superhorizon scales for any theory that is disformally
related to Horndeski's theory.  The difference between disformally
related curvature perturbations is found to be given in terms of the
comoving density perturbation associated with a single canonical scalar
field.  In General Relativity it is well-known that this quantity
vanishes on superhorizon scales through the Poisson equation that is
obtained on combining the Hamiltonian and momentum constraints, and we
confirm that a similar result holds for any theory that is disformally
related to Horndeski's scalar-tensor theory so long as the invertibility
condition for the disformal transformation is satisfied.  We also
consider the curvature perturbation at full nonlinear order in the
unitary gauge, and find that it is invariant under a general disformal
transformation if we assume that an attractor regime has been reached.
Finally, we also discuss the counting of degrees of freedom in theories
disformally related to Horndeski's.
\end{abstract}


\maketitle

\section{Introduction}

A primordial epoch of inflation and the late-time accelerated expansion
of the universe constitute two key elements of the standard model of
modern cosmology, which is in very good agreement with observational
data.  Many of the models proposed to try and explain these two epochs
of accelerated expansion rely on the introduction of an additional
scalar degree of freedom, either in the form of an unknown scalar field
in the matter sector, such as an inflaton or quintessence field, or as part of a
modified gravity sector, such as in $f(R)$ gravity or Brans-Dicke
scalar-tensor gravity.

Recently, efforts have been made to determine the most general form of
scalar-tensor theory that encompasses the examples mentioned above and
more.  In the spirit of effective field theories, such a theory would
allow one to introduce a common parameterisation for a wide range of
models, making it much easier to understand the relation between
different models and to compare their predictions with observations. 
In trying to construct the most general
form of scalar-tensor action, a key requirement is that the resulting
equations of motion are second order in time derivatives.  The appearance of higher order derivatives is generally associated with the presence of additional, ghost-like degrees of freedom, and the associated Hamiltonian
becomes unbounded from below, both in the presence of even higher order derivative terms \cite{Woodard:2006nt} and odd higher order derivative terms \cite{Motohashi:2014opa}.  As a result, if the system is coupled to
another ``normal'' system, then the total system will develop a
so-called Ostrogradsky instability.  The most general form of 
scalar-tensor action that gives rise to second order equations of
motion
was derived by Horndeski over 40 years ago
\cite{Horndeski:1974wa}, and was rederived just a few years ago in the
context of so-called Galileon models
\cite{Nicolis:2008in,Deffayet:2009wt,Deffayet:2011gz,Kobayashi:2011nu}.  
More recently, however, it has become apparent that there exist theories that do not
belong to Horndeski's theory but that nevertheless 
do not suffer from Ostrogradsky instabilities, propagating only 3 degrees of freedom
\cite{Zumalacarregui:2013pma,Gleyzes:2014dya,Gao:2014soa,Lin:2014jga,Gleyzes:2014qga,Kase:2014cwa,Gao:2014fra,Gleyzes:2014rba}.
Whilst
these theories may appear to give rise to higher-order equations of
motion at the level of the Euler-Lagrange equations -- which is why they
are not included in Horndeski's theory -- 
it has been shown that the higher-order time derivatives 
can be removed by making use of the time derivative of a special linear combination of the gravitational equations of motion,
thus rendering the equations of motion second order with respect to time derivatives \cite{Deffayet:2015qwa}.
These theories are therefore interesting and their phenomenology 
has been investigated in the literature, e.g.\
the screening mechanism \cite{Kobayashi:2014ida,Koyama:2015oma,Saito:2015fza}
and possible observational signatures \cite{DeFelice:2015isa}.

In exploring this class of general
scalar-tensor theories, use is often made of disformal transformations
of the metric, which take the form \cite{Bekenstein:1992pj}
\begin{equation}\label{disTran}
 \tilde g_{\mu\nu}=
\alpha(\phi,X)g_{\mu\nu} +
\beta(\phi,X)\partial_\mu\phi\partial_\nu\phi,
\end{equation}
where $X = -g^{\mu\nu}\partial_\mu\phi\partial_\nu\phi/2$. 
This is a generalisation of the more familiar conformal transformations,
for which $\beta(\phi,X) = 0$ and $\alpha(\phi,X)\rightarrow\alpha(\phi)$.  
The different representations of a
theory, written in terms of disformally related metrics, are often
referred to as being written in different ``frames''.  In some cases a
transformation of the form \eqref{disTran} can be used to remove
non-minimal coupling between the scalar field $\phi$ and the Ricci
scalar or Einstein tensor at the level of the action, leaving only a
canonical Einstein-Hilbert term
\cite{Maeda:1988ab,Bettoni:2013diz,Zumalacarregui:2013pma}.  This
particular frame, if it exists, is referred to as the Einstein
frame.  It is known that the form of Horndeski's action is preserved
under disformal transformations if $\alpha$ and $\beta$ only depend on
$\phi$ \cite{Bettoni:2013diz,Zumalacarregui:2013pma}.  
Similarly, the form of the theories beyond Horndeski's considered in \cite{Gleyzes:2014dya,Gleyzes:2014qga} is known to be preserved under disformal transformations with $\alpha = \alpha(\phi)$ and $\beta = \beta(\phi,X)$.
Allowing for an $X$-dependence of $\alpha$,
however, allows one to transform
between theories belonging to the class considered in \cite{Gleyzes:2014dya,Gleyzes:2014qga} 
and those that lie outside it~\cite{Zumalacarregui:2013pma,Gleyzes:2014qga}.

In many cases it is easier to solve for the dynamics of the scalar field
$\phi$ coupled to gravity if we first rewrite the theory in terms of a
metric that is disformally related to the original metric as in
\eqref{disTran}.  Having solved for $\phi$ and $\tilde g_{\mu\nu}$,
however, it is often the case that we would like to relate these
quantities back to the original metric $g_{\mu\nu}$, for example if
matter is minimally coupled to this metric, i.e.\ it defines the so-called Jordan frame.
In the context of cosmology
we are particularly interested in the transformation properties of
perturbations, and especially the so-called comoving curvature
perturbation, $\mathcal R_c$, which is defined as the curvature
perturbation on time-slices of constant $\phi$.  It has been known for
some time that the comoving curvature perturbation is invariant under
conformal transformations with $\beta(\phi,X) = 0$ and $\alpha(\phi,X)
\rightarrow \alpha(\phi)$, both at 
the linear level \cite{Makino:1991sg} and 
the fully nonlinear level \cite{Chiba:2008ia,Gong:2011qe}. 
More recently, the invariance of $\mathcal R_c$ was also confirmed for
transformations where $\alpha(\phi,X) \rightarrow \alpha(\phi)$ and
$\beta(\phi,X)\rightarrow \beta(\phi)$ \cite{Minamitsuji:2014waa}, and
finally for the case where $\alpha(\phi,X)\rightarrow\alpha(\phi)$ and
$\beta$ has both $\phi$- and $X$-dependence \cite{Tsujikawa:2014uza}.
The disformal invariance of the comoving curvature perturbation is a
very useful result, as it means that if we are ultimately only
interested in the comoving curvature perturbation then we are free to
solve the system in whichever frame is most convenient.
It is thus natural to ask whether or not the disformal
invariance of $\mathcal R_c$ holds for the most general form of
disformal transformation, where we allow for an $X$-dependence of both
$\alpha$ and $\beta$, and this is the question we address in this paper.
We will show that the case where an $X$-dependence of $\alpha$ is
included is crucially different to the previously considered cases, and
the comoving curvature perturbation is not identically invariant under
such a disformal transformation.

The paper is organised as follows. In \S \ref{sec:lin}, we investigate
the transformation properties of linear perturbations, choosing to leave
the gauge unfixed and to work with gauge-invariant quantities. 
We elucidate that the comoving curvature perturbation is not identically
invariant under disformal transformations when one allows for an
$X$-dependence of $\alpha$, but that the difference is given in terms of
the gauge-invariant comoving density perturbation associated with a
single canonical scalar field.  We then show that in the context of Horndeski's theory a sufficient condition for the comoving density
perturbation to vanish is $\dot{\mathcal R}_c=0$, and that under reasonable assumptions -- i.e.\ neglecting the so-called decaying mode of $\mathcal R_c$ -- this condition is satisfied on super-horizon scales.  Details of this calculation are
presented in Appendix~\ref{app:van}.  Consequently, we conclude that on
superhorizon scales, and under reasonable assumptions, the comoving curvature perturbation is disformally
invariant for any theory that is disformally related to Horndeski's
theory.  In \S \ref{sec:non}, we consider the transformation properties of
perturbations at the nonlinear level, and in this analysis we find it
easier to make use of the unitary gauge, where $\delta\phi=0$, which
makes $\beta$ irrelevant to the transformation law of the comoving
curvature perturbation.  We find that the comoving curvature
perturbation is invariant under disformal transformations if we assume
that an attractor regime has been reached.  In such an attractor regime
$X$ can be re-expressed as a function of $\phi$, meaning that the
situation is exactly the same as in the case of an $X$-independent
$\alpha$.  Using our result, we deduce that in the attractor regime the
nonlinear curvature perturbation is conserved on superhorizon scales in
any theory that is related to Horndeski's by a general disformal
transformation.  In \S \ref{sec:dishorn}, we discuss the counting of
degrees of freedom in theories disformally related to
Horndeski's.  Given that we know Horndeski's theory to be healthy,
i.e.\ its equations of motion are second order and it propagates only 3
degrees of freedom, we focus our attention on how the counting of
degrees of freedom is affected by a disformal transformation.  Using a
toy model we demonstrate that, as one would naively expect, the number of
degrees of freedom as determined by a full Hamiltonian analysis should
be unaffected by a disformal transformation, which is agreement with the recent analysis in \cite{Domenech:2015tca}.  \S \ref{sec:con} is devoted to conclusions.

\section{Linear analysis}
\label{sec:lin}

We consider the disformal transformation given in \eqref{disTran}.  To make it a well-defined redefinition of fields, we require that the inverse disformal transformation from the tilded frame to un-tilded frame also exists, which requires the existences of $\tilde g^{\mu\nu}$ and the solvability of $X=X(\phi,\tilde X)$.

Firstly, from \eqref{disTran} we can determine that the inverse metric $\tilde g^{\mu\nu}$ is given as 
\begin{equation}\label{invdg}
 \tilde g^{\mu\nu} = \frac{g^{\mu\nu}}{\alpha(\phi, X)} - \frac{\beta(\phi,X)\partial^\mu\phi\partial^\nu\phi}{\alpha(\phi,X)[\alpha(\phi,X)-2X\beta(\phi,X)]}, 
\end{equation}
where $\partial^\mu\phi = g^{\mu\nu}\partial_\nu\phi$.  
From this expression we see that the inverse $\tilde g^{\mu\nu}$ will exist as long as 
\be \alpha(\alpha - 2X\beta) \ne 0. \label{invtgcondition} \ee
This could also have been inferred from the relation between the determinants of $\tilde g_{\mu\nu}$ and $g_{\mu\nu}$, which can be derived by contracting \eqref{disTran} with $g^{\mu\nu}$ and taking the determinant \cite{Bekenstein:2004ne} 
\begin{equation}
 g^{\mu\nu}\tilde g_{\nu\alpha} = \alpha \left(\delta^\mu{}_\alpha+ \frac{\beta}{\alpha}\partial^\mu\phi\partial_\alpha\phi\right)\qquad\Rightarrow\qquad \frac{\tilde g}{g} = \alpha^3(\alpha - 2X\beta),
\end{equation}
where $\tilde g$ and $g$ are the determinants of $\tilde g_{\mu\nu}$ and $g_{\mu\nu}$ respectively.

Secondly, let us consider the
solvability condition for $X$ in terms of $\phi$ and $\tilde X$.
The complication that arises in trying to invert the transformation given in \eqref{disTran} is the appearance of $g^{\mu\nu}$ in $X$.  As discussed in \cite{Arroja:2015wpa}, in order to be able to invert the transformation we 
thus need to express $X=-g^{\mu\nu}\partial_\mu\phi\partial_\nu\phi/2$ in terms of $\tilde X = -\tilde g^{\mu\nu}\partial_\mu\phi\partial_\nu\phi/2$.
When \eqref{invtgcondition} is satisfied, we can contract \eqref{invdg} with $\partial_\mu\phi\partial_\nu\phi$ and
obtain the relation 
\begin{equation}\label{tildex}
 \tilde X = \frac{X}{\alpha(\phi,X) - 2X\beta(\phi,X) }.
\end{equation}
The solubility of this relation for $X$
requires $\partial \tilde X/\partial X \neq 0$, namely,
\begin{equation}\label{invcondition}
 \f{\alpha-X\alpha_X+2X^2\beta_X}{(\alpha-2X\beta)^2}\neq 0,
\end{equation}
where $\alpha_X = \partial_X\alpha$ and $\beta_X = \partial_X \beta$.\footnote{We will adopt a similar notation for derivatives with respect to $\phi$, e.g.\ $\alpha_\phi = \partial_\phi \alpha$.}
The same condition can also be determined by considering the Jacobian $\partial \tilde g_{\mu\nu}/\partial g_{\alpha\beta}$ and requiring its determinant to be non-vanishing, as in~\cite{Zumalacarregui:2013pma}.
Combined with \eqref{invtgcondition},
\be \alpha(\alpha - 2X\beta)(\alpha-X\alpha_X+2X^2\beta_X) \ne 0 \label{discond} \ee
is the necessary and sufficient condition for the invertibility of the disformal transformation.  
If \eqref{discond} is satisfied, we can solve \eqref{tildex} for $X=X(\phi,\tilde X)$ and obtain the inverse disformal transformation
\be  g_{\mu\nu} = \frac{1}{\alpha(\phi,X(\phi,\tilde X))} \tilde g_{\mu\nu} - \frac{\beta(\phi,X(\phi,\tilde X))}{\alpha(\phi, X(\phi,\tilde X))} \partial_\mu \phi \partial_\nu \phi. \ee 
In the following, we consider disformal transformations that satisfy the invertibility condition \eqref{discond}.

We first focus on a linear analysis of perturbations.  Let us take a
line element of the form:
\begin{align}\label{linPert}
 \nonumber ds^2 &\equiv g_{\mu\nu} dx^\mu dx^\nu \\ \nonumber &=
 -(1+2A)dt^2 + 2a(\partial_iB-S_i)dx^idt \\ & \qquad\quad + a^2 [
 (1-2\psi)\delta_{ij}+2\partial_i\partial_j E + \partial_iF_j +
 \partial_jF_i+h_{ij} ] dx^idx^j ,
\end{align}
where $F_i$ and $h_{ij}$ satisfy $F_{i,i}=h_{ii}=h_{ij,i}=0$.\footnote{Note that repeated indices are summed over.} We
then consider a disformal transformation of the form
\eqref{disTran}.
\begin{align}
\nonumber d\tilde s^2 &\equiv \tilde g_{\mu\nu} dx^\mu dx^\nu \\
\nonumber &= [-\alpha_0
(1+2A)-\delta\alpha+(\beta_0+\delta\beta)\dot\phi^2 + 2\beta_0\dot\phi
\dot{\delta\phi} ] dt^2 \\ \nonumber &\quad + 2[\alpha_0 a(\partial_i
B-S_i) + \beta_0\dot\phi\partial_i\delta\phi ]dt dx^i \\ &\quad +
\alpha_0 a^2 \kk{ \mk{1-2\psi + \f{\delta\alpha}{\alpha_0}} \delta_{ij}
+ 2\partial_i \partial_j E+\partial_iF_j+ \partial_jF_i+h_{ij} } dx^idx^j,
\end{align}
where a dot denotes $d/dt$ and we have decomposed $\alpha = \alpha_0 +
\delta\alpha$ and $\beta = \beta_0 + \delta\beta$.  
Note that we are
choosing to leave the gauge unfixed, which will make the interpretation later on more obvious.
We would then like
to rewrite this new line element in the same form as \eqref{linPert},
namely as
\begin{align}\nonumber
 d\tilde s^2 &= -(1+2\tilde A)d\tilde t^2 + 2\tilde a (\partial_i\tilde
 B-\tilde S_i)dx^id\tilde t \\ &\qquad\quad+ \tilde a^2 [ (1-2\tilde
 \psi)\delta_{ij}+2\partial_i \partial_j \tilde E + \partial_i\tilde F_j +
 \partial_j\tilde F_i+\tilde h_{ij} ]dx^idx^j .
\end{align}
At background level this then gives us
\begin{equation}\label{bgRel}
 \tilde a = a\sqrt\alpha_0\qquad\mbox{and}\qquad d\tilde t =
 \sqrt{\alpha_0 - \beta_0\dot\phi^2} dt ,
\end{equation}
where we note that a dot still corresponds to taking the derivative with
respect to $t$ rather than $\tilde t$.  At the level of perturbations,
from the $00$-component we obtain
\begin{equation} \label{tA}
 \tilde A = \frac{1}{\alpha_0 - \beta_0\dot\phi^2} \mk{ \alpha_0 A +
 \f{1}{2} \delta\alpha - \f{1}{2} \dot\phi^2 \delta\beta - \beta_0
 \dot\phi \dot{\delta\phi} }.
\end{equation}
From the $0i$-component we have
\begin{equation}
 \tilde B = \sqrt{\f{\alpha_0}{\alpha_0-\beta_0\dot\phi^2}}  \left(
 B + \frac{\beta_0 \dot\phi }{\alpha_0 a} \delta\phi \right)
 \qquad\mbox{and}\qquad \tilde S_i = S_i.
\end{equation}
Finally, from the $ij$-component we find
\begin{equation}\label{Rrel}
 \tilde \psi = \psi - \frac{\delta\alpha}{2\alpha_0},
\end{equation}
whilst $E$, $F_i$, and $h_{ij}$ remain unchanged.  As such, we see that the vector
and tensor perturbation are invariant under a general disformal
transformation at linear level.  
Whilst the above expressions are problematic if $\alpha_0 - \beta_0\dot\phi^2$ vanishes, recall that we are assuming $\alpha - 2X\beta  \neq 0$ for the existence of $\tilde g^{\mu\nu}$, which yields $\alpha_0 - \beta_0\dot\phi^2 \neq 0$ at background level.\footnote{In fact, the requirement that the perturbative expansion be valid in the disformally related frame, e.g. $\tilde A\ll 1$, will put tighter constraints on the background-dependent coefficients appearing in eqs. \eqref{tA}--\eqref{Rrel} than those imposed by the invertibility of the transformation, but we will not consider this issue here.}

Turning to the gauge-invariant comoving curvature perturbation, $\R$, it
is defined in the original frame as
\begin{equation}
 \R = - \psi - \frac{H}{\dot\phi}\delta\phi,
\end{equation}
where $H = \dot a /a$. In the new frame we similarly have
\begin{equation}
 \tR = - \tilde \psi - \frac{\tilde{H}}{d\phi/d\tilde t}\delta\phi,
\end{equation}
where $\tilde{H} = (1/\tilde a)d\tilde a/d\tilde t$.  Using the
background relations \eqref{bgRel} we have
\begin{equation} \label{dpdt}
 \frac{d\phi}{d\tilde t} = \frac{\dot\phi}{
 \sqrt{\alpha_0-\beta_0\dot\phi^2} } \qquad\mbox{and}\qquad \tilde{H} =
 \frac{1}{ \sqrt{\alpha_0-\beta_0\dot\phi^2} } \left( H +
 \frac{\dot\alpha_0}{2\alpha_0} \right),
\end{equation}
which, on combining with \eqref{Rrel}, gives us
\begin{equation}
 \tR = \R +
 \frac{1}{2\alpha_0}\left(\delta\alpha-\frac{\dot\alpha_0}{\dot\phi}\delta\phi
 \right).
\end{equation}
In the case that $\alpha = \alpha(\phi)$ 
we have $\delta\alpha - \dot\alpha_0\delta\phi/\dot\phi = 0$, 
meaning that $ \R = \tR$,
which is consistent with \cite{Minamitsuji:2014waa,Tsujikawa:2014uza}.
However, in the case
that we allow for an $X$-dependence of $\alpha$, we more generally get
\begin{equation}
 \delta\alpha -\frac{\dot \alpha_0}{\dot \phi}\delta\phi = \alpha_{0X}
 \left(\delta X - \frac{\dot X_0}{\dot \phi}\delta\phi \right),
\end{equation}
where $\alpha_{0X} = \partial\alpha_0/\partial X$ and $X_0 = \dot \phi^2/2$.
Using $\delta X = \dot \phi ( \dot{\delta\phi} - \dot \phi A )$,
we thus find
\begin{equation}\label{zetaDiff}
\tR - \R = \frac{\alpha_{0X}}{2\alpha_0}\Sp,
\end{equation}
where
\begin{equation}\label{comdp}
 \Sp \equiv \dot\phi (\dot{\delta\phi} - \dot\phi A) - \ddot \phi \delta\phi = \delta X - \ddot\phi\delta\phi.
\end{equation}
We thus see that the comoving curvature perturbation 
is not identically invariant under disformal transformations with
$\alpha=\alpha(\phi,X)$.  Note that if one takes the gauge $\delta\phi =
0$, $\tR$ as determined by \eqref{zetaDiff} coincides with the quantity
$\zeta_{\rm new}$ defined in Eq.~(96) of
\cite{Gleyzes:2014rba}.\footnote{In fact, the two expressions do not
exactly coincide, but this is due to a typo in Eq.~(96)
of \cite{Gleyzes:2014rba}, where $\tilde N$ should be
replaced by $N(=1+\delta N)$, so that to linear order the expression for $\zeta_{\rm new}$ should be $\zeta_{\rm new}=\zeta+\f{\Omega_N}{\Omega}\delta N$.
We thank J.\ Gleyzes for confirming this point.}  
The importance of $\zeta_{{\rm new}}$ was discussed in
\cite{Gleyzes:2014rba}: it absorbs all terms in the action generated by
a disformal transformation that explicitly depend on the time derivative
of the perturbation of the lapse function.  In so doing, it makes it
explicitly clear that no additional degrees of freedom appear as a
result of the disformal transformation.  In light of the above analysis,
we see that the appearance of the quantity $\zeta_{\rm new}(=\tR)$ is in
fact very natural, as it simply corresponds to the transformed comoving
curvature perturbation.

The quantity $\Sp$ is a gauge-invariant quantity corresponding to the gauge-invariant perturbation of $X$, as seen in \eqref{comdp}.
It also coincides with the comoving
density perturbation for a single canonical scalar field, 
$\Sp = \delta\rho_s =\delta\rho-3H\delta q$,
where $\delta \rho = \delta X + V_\phi \delta\phi$ is the density perturbation and  
$\delta q=-\dot\phi\delta\phi$ is the velocity potential for the energy
momentum tensor of the scalar field \cite{Gordon:2000hv}.
$\Sp$ is
also related to the intrinsic entropy perturbation of a canonical
scalar field as
\be \mathcal S \equiv H \mk{\f{\delta p}{\dot p}-\f{\delta\rho}{\dot\rho}} =
\f{2V_\phi}{3\dot\phi^2(3H\dot\phi+2V_\phi)} \Sp. 
\ee

So far, we have not assumed any particular scalar-tensor theory, and thus
\eqref{zetaDiff} holds for any theory.  We now proceed to consider
specific theories, in order to determine how $\Sp$ behaves.  In General
Relativity, if the scalar field is the dominant energy component of the
universe, then from Einstein's equations we are able to determine that
$\Sp$ satisfies the Poisson equation
\begin{equation}
-\f{k^2}{a^2}\Psi= \f{\Sp}{2}, \label{GRcase}
\end{equation}
where $\Psi\equiv \psi+a^2H(\dot
E-B/a)$ is the gauge-invariant Bardeen potential.  As such, $\Sp$ is suppressed
by $k^2$ on large scales as long as $\Psi$ remains finite, which implies that the difference between
$\tR$ and $ \R$ will also vanish on large scales.

It has also been shown that in a subclass of Horndeski's scalar-tensor
theory $\Sp$ still vanishes on superhorizon scales if the comoving
curvature perturbation remains constant
\cite{Minamitsuji:2014waa}. To
the best of our knowledge, however, it has not yet been explicitly shown
for the full Horndeski theory, and this is what we will now proceed to
confirm.  Here we simply give the result, and more details can be found
in Appendix~\ref{app:van}.

Horndeski's action takes the form
\begin{equation}\label{hornAct}
 S = \int d^4x\sqrt{-g} \mathcal L_H,
\end{equation}
with $\mathcal L_H=\sum_{i=2}^5 \mathcal L_i$ and 
\begin{align}
\mathcal L_2 &= K(\phi,X),\\
\mathcal L_3 &= -G_3(\phi,X)\Box\phi,\\
\mathcal L_4 &= G_4(\phi,X) R +
 G_{4X}\left[ (\Box\phi)^2 - ( \nabla_\mu \nabla_\nu \phi) ( \nabla^\mu \nabla^\nu \phi ) \right],\\
\mathcal L_5 &= G_5(\phi,X)G_{\mu\nu}\nabla^\mu\nabla^\nu\phi -
 \frac{1}{6}G_{5X}\left[ (\Box\phi)^3 - 3(\Box \phi)(\nabla_\mu\nabla_\nu\phi)(\nabla^\mu\nabla^\nu\phi)+2(\nabla^\mu\nabla_\alpha\phi)(\nabla^\alpha\nabla_\beta\phi)(\nabla^\beta\nabla_\mu\phi) \right].
\end{align}
Focusing on scalar perturbations, and taking the spatial gauge $E=0$ in \eqref{linPert}, the
equations of motion for $A$, $B$, $\psi$ and $\delta\phi$ were derived
in \cite{DeFelice:2011hq}.  Combining the constraint equations that
result from varying the second order action with respect to $A$ and $B$,
we are able to derive the following Poisson equation for $\Sp$
\begin{align}\label{genPoiss}
\Sp =
 \frac{k^2}{a^2H^2}\frac{\dot\phi^2 C_1 H^2}{C_1A_4-C_3A_1}\left(A_3\mathcal R_c + \frac{A_5}{H}(\Psi + \mathcal
 R_c)\right),
\end{align}
where the coefficients $A_i$ and $C_i$ depend only on
background quantities and are given in Appendix~\ref{app:van}.  In the case of General Relativity with a single canonical scalar field, 
where $G_3 = G_5 = 0$, $K(\phi,X) = X - V(\phi)$ and $G_4 =
1/2$, we have
\begin{equation}\label{GRCoeff}
 A_1 = 6H,\qquad A_3 = 2,\qquad A_4 = 2X-6H^2,\qquad A_5 =
-2H,\qquad C_1 = 2\qquad \mbox{and} \qquad C_3 = -2H, 
\end{equation}
so that \eqref{genPoiss} reduces to \eqref{GRcase}.

Given the form of \eqref{genPoiss}, one can conclude that $\Sp$ vanishes on superhorizon scales --- i.e.\ in the limit $k\ll aH$ --- so long as the coefficient of $k^2/(aH)^2$ on the right hand side is finite in this limit.  Alternatively, from the momentum constraint we have 
\begin{equation}
\Sp = \frac{C_1}{C_3}\dot\phi^2\dot{\mathcal R}_c, \label{esrc}
\end{equation}
which is given in \eqref{gInvMomCon} but we repeat here for convenience.  As such, we see that a sufficient condition for the vanishing of $\Sp$ -- and thus disformal invariance of $\mathcal R_c$ -- is that $\mathcal R_c$ is conserved.

In the case of General Relativity plus canonical scalar field, it is well known that $\R$ is conserved on superhorizon scales, provided that the so-called decaying mode can be neglected.  Explicitly, one finds that $\dot{\mathcal R}_c \propto 1/(\epsilon a^3)$, where $\epsilon = \dot\phi^2/(2H^2)$.  As such, we see that $\dot{\mathcal R}_c$ is indeed decaying --- and therefore negligible --- provided $\epsilon$ is not decaying faster than $a^{-3}$.  This condition is satisfied in almost all standard slow-roll inflation models, where the slow-roll parameter $\epsilon$ is itself taken to be slowly varying.  There is, however, a special class of inflation models -- dubbed ``ultra-slow-roll'' models -- for which extra care is needed~\cite{Kinney:2005vj,Namjoo:2012aa,Martin:2012pe,Motohashi:2014ppa}.  In the simplest ultra-slow-roll inflation model with constant potential one finds $\epsilon \propto a^{-6}$, which means that $\dot{\mathcal R}_c$ is growing as $a^3$.  Interestingly, however, we still find that $\Sp$ decays as $a^{-3}$, which follows from the fact that $\Sp$ and $\dot{\mathcal R}_c$ are related by a factor of $\dot\phi^2\propto a^{-6}$.  As such, provided the disformal transformation is such that the factor $\alpha_{0X}/\alpha_0$ appearing in \eqref{zetaDiff} is not growing faster than $a^3$, we see that $\R$ is disformally invariant even in the case of ultra-slow-roll inflation.

In the more general case of Horndeski's theory we have a similar result.  As was shown in \cite{Kobayashi:2011nu}, on superhorizon scales $\R$ has the two independent solutions\footnote{Note that these two solutions are in fact valid on scales larger than the sound horizon of the scalar perturbation $\R$.  Horizon crossing is defined by $c_s^2k^2 = a^2H^2$, where $c_s$ is the sound speed of $\R$ and is in general different from unity.  See \cite{Kobayashi:2011nu} for the general expression.} 
\begin{equation}
 \R = {\rm const.}\qquad{\rm and}\qquad \R \propto \int^t\frac{1}{\mathcal G_Sa^3}dt^\prime,
\end{equation}    
where $\mathcal G_S = (\Sigma/\Theta^2)\mathcal G_T^2 + 3\mathcal G_T$ and $\Sigma$, $\Theta$ and $\mathcal G_T$ are as defined in Appendix~\ref{app:van}.  We thus find that $\dot{\mathcal R}_c \propto 1/(\mathcal G_S a^3)$, which is decaying provided $\mathcal G_S$ is not decaying faster than $a^{-3}$.  In the case of standard slow-roll inflation we expect this to be the case, but there will be exceptions analogous to ultra-slow-roll inflation.  Strictly speaking, even if the decaying mode can be neglected, in order to then conclude that $\tilde{\mathcal R}_c = \R$ on superhorizon scales we additionally must assume that the combination $\alpha_{0X}C_1\dot\phi^2/(\alpha_0 C_3)$ is not growing faster than $\dot{\mathcal R}_c$ is decaying.  This again seems reasonable if we assume that background quantities are evolving slowly, but perhaps there may be some exceptions in the very general context of Horndeski's theory.  Conversely, as was the case with ultra-slow-roll inflation, even if $\dot{\mathcal R}_c$ is not decaying, $\R$ may still be disformally invariant if the combination $\alpha_{0X}C_1\dot\phi^2/(\alpha_0 C_3)$ is decaying faster than $\dot{\mathcal R}_c$ is growing.

To reiterate, our main conclusion is that in the context of Horndeski's theory, a sufficient condition for the vanishing of $\Sp$ -- and thus disformal invariance of $\R$ -- is that $\R$ is conserved, and this is the case on superhorizon scales so long as we can neglect the so-called decaying mode of $\R$.  Models in which the decaying mode cannot be neglected -- such as ultra-slow-roll inflation --  must be considered on a case-by-case basis, but interestingly it seems that the disformal invariance of $\R$ does not necessarily break down in such cases.  For the remainder of this section we will restrict ourselves to considering models in which the decaying mode can be neglected.

Using the above results, we can argue that the comoving density perturbation $\Sp$ should
vanish on superhorizon scales in any theory that is disformally related
to Horndeski's theory as follows.  Suppose we have two theories, theory
A and theory B, that are both disformally related to an element of
Horndesdki's theory, theory H.  The metrics for these theories are
related as
\begin{align}
\gmn^\kA &= \alpha^\kHA \gmn^\kH + \beta^\kHA \pmn, \\
\gmn^\kB &= \alpha^\kHB \gmn^\kH + \beta^\kHB \pmn. 
\end{align}
We can then consider a disformal transformation between theory A and
theory B 
\be \gmn^\kB = \alpha^\kAB \gmn^\kA + \beta^\kAB \pmn, \ee
with 
\begin{align}
\alpha^\kAB &= \f{ \alpha^\kHB }{ \alpha^\kHA } , \\
\beta^\kAB &= \beta^\kHB - \f{ \alpha^\kHB }{ \alpha^\kHA } \beta^\kHA .
\end{align}
Then, the comoving curvature perturbations in these theories are related as
\begin{align}
\R^\kA - \R^\kH &= \f{\alpha_{0X}^\kHA}{2\alpha_0^\kHA} \Sp^\kH, \\
\R^\kB - \R^\kH &= \f{\alpha_{0X}^\kHB}{2\alpha_0^\kHB} \Sp^\kH.
\end{align}
Here, $\alpha_{0X}^{({\rm P}\to{\rm Q})}$ is understood as a derivative with respect to $X^{({\rm P})}\equiv g_{({\rm P})}^{\mu\nu}\pmn$.
As the comoving density perturbation $\Sp^\kH$ vanishes on superhorizon scales in Horndeski's theory, 
$\R^\kA = \R^\kB = \R^\kH $ on superhorizon scales.
By considering the disformal transformation of $\R$ between theories A and
B, which gives us 
\be \R^\kB - \R^\kA = \f{\alpha_{0X}^{\kAB}}{2\alpha_0^{\kAB}} \Sp^\kA, \ee
the vanishing of the left hand side on superhorizon scales allows us to
infer the vanishing of $\Sp^\kA$ on superhorizon scales.

Indeed, we can also confirm the above statement explicitly by
considering how $\Sp$ transforms under a disformal transformation.
Using \eqref{tA}, \eqref{dpdt} and \eqref{comdp}, we obtain
\be \label{tsp}  
\tSp = 
\f{\alpha_0 - \alpha_{0X} \f{\dot\phi^2}{2} + 2 \beta_{0X} \mk{\f{\dot\phi^2}{2}}^2 }{(\alpha_0-\beta_0\dot\phi^2)^2}
\Sp, 
\ee
which recovers the result in \cite{Minamitsuji:2014waa} when $\alpha$
and $\beta$ are functions of $\phi$ only. Interestingly, even for
a general disformal transformation with $X$-dependent $\alpha$ and $\beta$, $\Sp$ is
disformally invariant up to a coefficient depending on background
quantities.  This implies that any theory disformally related to Horndeski's theory should also have vanishing $\Sp$ on large scales.

Let us recall here that we are assuming the invertibility condition \eqref{discond} is satisfied.  This condition precisely guarantees that the coefficient in front of $\epsilon_s$ in \eqref{tsp} neither diverges nor vanishes.  As such, the conclusion that $\tilde \epsilon_s$ vanishes if $\epsilon_s$ vanishes holds for any disformal transformation that is invertible.  It is interesting to note that the coefficient on the right-hand side of \eqref{tsp} exactly coincides with the left-hand side of \eqref{invcondition} evaluated at background level, i.e.\ it coincides with $\partial\tilde X/\partial X$.  This can be expected, however, as $\tilde\epsilon_s$ corresponds to the gauge-invariant perturbation of $\tilde X$ and $\Sp$ to the gauge-invariant perturbation of $X$ (recall \eqref{comdp}).

\section{Nonlinear analysis}
\label{sec:non}

In going beyond linear perturbations let us take the unitary gauge from
the outset, where $\delta\phi = 0$.  In this case, the spatial part of
the metric is not affected by the disformal part of the
disformal transformation, $\beta$, due to the fact that $\partial_i\phi = 0$.
Let us take the metric with nonlinear perturbations as
\begin{equation}
ds^2 = -N^2 dt^2 +
a^2e^{2\R}\gamma_{ij}\left(dx^i+N^idt\right)\left(dx^j + N^i
dt\right),
\end{equation}
where \be \gamma_{ij}=e^{2\partial_i\partial_j E + \partial_iF_j +
\partial_jF_i+h_{ij}} \ee and $F_i$ and $h_{ij}$ once again satisfy
$F_{i,i}=h_{ii}=h_{ij,i}=0$. Note also that $N^i$ contains both scalar and vector components, and that spatial indices should be raised and lowered with $\gamma^{ij}$ and $\gamma_{ij}$, respectively.  If we further parameterise
$\alpha(\phi,X)$ as $\alpha \equiv \alpha_0 e^{2\Delta\alpha}$, where
$\alpha_0$ corresponds to the background part of $\alpha$ and
$e^{2\Delta\alpha}$ contains nonlinear perturbations from this background
value, then we see that in the gauge $\delta\phi = 0$ the
metric transforms as
\begin{align}
 \tilde g_{00} &= -\alpha N^2 +\beta\dot\phi^2 + \alpha_0 a^2 e^{2\R +
 2\Delta\alpha} \gamma_{ij} N^i N^j, \nonumber \\ \tilde 
 g_{0i} &= \alpha_0 a^2 e^{2\R + 2\Delta\alpha} \gamma_{ij} N^j, \nonumber \\
 \tilde g_{ij} &= \alpha_0 a^2 e^{2\R + 2\Delta\alpha} \gamma_{ij},
\end{align}
from which we deduce
\begin{equation}\label{nonLinTran}
 \tilde N^2= \alpha N^2 -\beta\dot\phi^2, \qquad \tilde a^2 = \alpha_0
 a^2, \qquad \tR = \R + \Delta\alpha,
\end{equation}
whilst $E$, $F_i$, $N^i$ and $h_{ij}$ remain unchanged at nonlinear
level, which is consistent with the result at linear level in the
previous section.  As such, the vector and tensor perturbations are
invariant at the nonlinear level.

In analysing the last relation in \eqref{nonLinTran}, note that as
$\beta$ is irrelevant for the transformation law of $\R$, the situation
is equivalent to determining how $\R$ transforms under a conformal
transformation.  As such, in the case that $\alpha = \alpha(\phi)$ our
conclusion is the same as that reached in \cite{Gong:2011qe}: in the
unitary gauge $\delta\phi = 0$ so that $\Delta\alpha = 0$, meaning that
$\R$ is invariant.  This is also in agreement with
\cite{Tsujikawa:2014uza} and -- at the linear level -- with the results
of \S \ref{sec:lin} and \cite{Minamitsuji:2014waa}.

In contrast, when we allow for an $X$-dependence of $\alpha$, even in
the unitary gauge we find that $\Delta\alpha \ne 0$ as a result of the
dependence of $X$ on $N$.  Explicitly, in the unitary gauge we have 
$\alpha = \alpha(\phi,\dot\phi^2/(2N^2))$.
As such, we see that $\alpha$ will only coincide with its background
value, hence giving $\Delta\alpha = 0$ and $\tR = \R$, when the
perturbation of the lapse function vanishes.  At linear order, this
condition corresponds to $A=0$, which is thus consistent with the 
requirement found in \S \ref{sec:lin} that $\Sp$ must vanish
if we are to have $\tR = \R$, as in the unitary gauge we have $\Sp =
-\dot\phi^2 A$.

In order to aid an intuitive understanding of this condition, let us
define the proper time $\tau$ as $d\tau = N dt$.  Starting with the
definition of $\Sp$ at linear order, we can see that it can be
rewritten as
\begin{equation}
 \Sp = \partial_\tau\phi\partial^2_\tau\phi\left(\frac{\delta(\partial_\tau\phi)}{\partial^2_\tau\phi} - \frac{\delta\phi}{\partial_\tau\phi}\right),
\end{equation}
where the term in brackets corresponds to the relative entropy
 perturbation between $\partial_\tau\phi$ and $\phi$.  We can thus see  
that $\Sp$ vanishes -- in turn giving $\tR = \R$ -- when $\partial_\tau\phi = f(\phi)$, where $f(\phi)$
 is some function of $\phi$.  Similarly, turning to the nonlinear
 case, we see that in terms of $\tau$ we can write $\alpha =
\alpha(\phi,(\partial_\tau\phi)^2/2)$.  Imposing 
$\partial_\tau\phi = f(\phi)$ means that in the unitary gauge $\alpha$
 is equal to its background value, which in turn gives us $\Delta\alpha
 = 0$ and $\tR=\R$.  The condition $\partial_\tau\phi = f(\phi)$ is
 familiar to us as the condition for an attractor regime (see e.g.\
\cite{Naruko:2011zk}), and we thus conclude that in the unitary gauge and an
attractor regime the curvature perturbation is disformally invariant at
the nonlinear level.  Note that the requirement to be in an attractor
regime is not as restrictive as it may sound.  Indeed, the vast majority
of standard inflationary models satisfy this condition.

It has been shown in \cite{Gao:2011mz} that even at the nonlinear level the comoving
curvature perturbation has a mode that remains constant on superhorizon scales in
Horndeski's theory.  Provided that the other so-called decaying mode can be
neglected, this allows us to conclude that on superhorizon scales the nonlinear comoving curvature perturbation is both conserved
and disformally invariant in any theory that is disformally related to
Horndeski's if one is in the attractor regime.

\section{The number of degrees of freedom in theories disformally related to Horndeski's}
\label{sec:dishorn}

In the preceding sections we have investigated the transformation
properties of the comoving curvature perturbation both at the linear and
nonlinear level, with our conclusions holding for any scalar-tensor
theory that is disformally related to Horndeski's theory.  In this
section we discuss whether or not such theories are well behaved in the
sense that they do not suffer from so-called Ostrogradsky instabilities.
Such instabilities generically arise in theories possessing equations of
motion that are higher than second order in time derivatives, indicating
the presence of additional ghost-like degrees of freedom.  As we will
see below, the equations of motion derived from theories disformally
related to Horndeski's theory do seemingly contain higher-order derivatives, thus suggesting that they are unhealthy.  However, this is
somewhat at odds with our expectation given that the theories are simply
related to instability-free Horndeski's theory by a redefinition of
fields.  In the following we try to address this apparent contradiction.

The action associated with the aforementioned class of theories can be
written in the following form \be \label{genact}S = \int d^4x \sqrt{-\tilde g}
\mathcal{L}_H(\tilde g_{\mu\nu},\phi) + \int d^4x \sqrt{- g}
\mathcal{L}_m(g_{\mu\nu}),\ee where we have Horndeski's Lagrangian,
$\mathcal L_H$, written in terms of the metric $\tilde g_{\mu\nu}$ and
matter is minimally coupled to the metric $g_{\mu\nu}$.  The metric
$\tilde g_{\mu\nu}$ is disformally related to $g_{\mu\nu}$ as in
\eqref{disTran}.  The frame defined by $\tilde g_{\mu\nu}$ is the
``Horndeski frame'', in which the gravitational Lagrangian coincides
with that of Horndeski's theory and matter is non-minimally coupled to
the scalar field through $g_{\mu\nu}=(\tilde
g_{\mu\nu}-\beta\partial_\mu \phi \partial_\nu \phi)/\alpha$.  However,
we are interested in the equations of motion in the Jordan frame,
defined by $g_{\mu\nu}$, in which matter is not coupled to the scalar
field.  Varying the above action with respect to $g_{\mu\nu}$ and $\phi$
yields their equations of motion,
\begin{align}
\label{eq1} &\alpha \mathcal{E}^{\mu\nu}_H + \f{1}{2} \mathcal{E}^{\rho\sigma}_H \mk{ \alpha_X g_{\rho\sigma} + \beta_X \partial_\rho \phi \partial_\sigma \phi } \partial^\mu \phi \partial^\nu \phi + \f{1}{2\alpha \sqrt{\alpha(\alpha-2X\beta)}} T^{\mu\nu}_m = 0, \\
\label{eq2} &\nabla_\mu \kk{ \alpha \sqrt{\alpha(\alpha-2X\beta)}\ck{\mathcal{E}^{\rho\sigma}_H \mk{ \alpha_X g_{\rho\sigma} + \beta_X \partial_\rho \phi \partial_\sigma \phi } \partial^\mu \phi - 2\beta \mathcal{E}^{\mu\nu}_H \partial_\nu \phi}} \notag\\ &\hspace{2.3cm}+\alpha \sqrt{\alpha(\alpha-2X\beta)}\kk{\mathcal{E}^{\rho\sigma}_H \mk{ \alpha_\phi g_{\rho\sigma} + \beta_\phi \partial_\rho \phi \partial_\sigma \phi } + \mathcal{E}^{(\phi)}_H} = 0, 
\end{align}
where $\mathcal{E}^{\mu\nu}_H$ and $\mathcal{E}^{(\phi)}$ are determined
by varying Horndeski's Lagrangian with respect to $\tilde g_{\mu\nu}$
and $\phi$ \cite{Kobayashi:2011nu,Gao:2011mz}, namely, $\delta
(\sqrt{-\tilde g} \mathcal{L}_H) = \sqrt{-\tilde g}
(\mathcal{E}^{\mu\nu}_H \delta \tilde g_{\mu\nu} + \mathcal{E}^{(\phi)}
\delta \phi)$, and $T^{\mu\nu}_m = \f{2}{\sqrt{-g}} \f{\delta(\sqrt{-g}
\mathcal{L}_m)}{\delta g_{\mu\nu}}$ denotes the energy-momentum tensor
associated with the matter Lagrangian.  
Note that here $\partial^\mu\phi = g^{\mu\alpha}\partial_\alpha\phi$.  
As $\mathcal{E}^{\mu\nu}_H$ and
$\mathcal{E}^{(\phi)}$ are the equations of motion for Horndeski's
theory, they contain at most second order derivatives of $\phi$ and
$\tilde g_{\mu\nu}$, meaning that they also contain at most second order
derivatives of $g_{\mu\nu}$. However, as $\tilde g_{\mu\nu}$ contains
$X = -g^{\mu\nu}\partial_\mu\phi\partial_\nu\phi/2$, 
in principle it is possible for $\mathcal{E}^{\mu\nu}_H$ and
$\mathcal{E}^{(\phi)}$ to contain up to third order derivatives of
$\phi$.  As such, we see that \eqref{eq1} will contain up to second
order derivatives of $g_{\mu\nu}$ but potentially third order
derivatives of $\phi$.  Similarly, \eqref{eq2} will contain up to third
order derivatives of $g_{\mu\nu}$ and fourth order derivatives of
$\phi$.

As already mentioned, it is reasonable to expect that the appearance of
seemingly dangerous higher order derivatives in \eqref{eq1} and
\eqref{eq2} may be spurious, as we know that our theory is related to a
healthy theory by a field redefinition.  Indeed, following the
analysis of \cite{Deruelle:2014zza,Arroja:2015wpa}, it is possible to
show that the gravitational equations of motion \eqref{eq1} are
equivalent to the equations of motion obtained by varying the action with respect to $\tilde g_{\mu\nu}$ so long as the transformation \eqref{disTran} is invertible.  Explicitly, we obtain (see Appendix~\ref{app:equiv} for details)
\be \label{eq1s}
\mathcal{E}^{\mu\nu}_H +\f{1}{2}T^{\mu\nu}_H = 0,
\ee
where $T^{\mu\nu}_H = \f{2}{\sqrt{-\tilde g}} \f{\delta(\sqrt{-g}
\mathcal{L}_m)}{\delta \tilde g_{\mu\nu}}$ and is given explicitly in terms of $T^{\mu\nu}_m$ as
\begin{equation}\label{Tmunurel}
  T^{\mu\nu}_H = \f{1}{\alpha^2 \sqrt{\alpha(\alpha-2X\beta)}}
\kk{ T^{\mu\nu}_m - \f{\partial^\mu \phi \partial^\nu
\phi}{2(\alpha+2X^2\beta_X-X\alpha_X)} T^{\rho\sigma}_m \mk{\alpha_X
g_{\rho\sigma} + \beta_X \partial_\rho \phi \partial_\sigma \phi } }.
\end{equation}
Note that $T^{\mu\nu}_H$ contains at most first order
derivatives of $\phi$.  On substituting the above expression for
$\mathcal{E}^{\mu\nu}_H$ into \eqref{eq2}, we obtain
\begin{align} \label{eq2s} 
&\alpha \sqrt{\alpha(\alpha-2X\beta)}~ \mathcal{E}^{(\phi)}_H  + \nabla_\mu \kk{ \f{\beta}{\alpha} T^{\mu\nu}_m \partial_\nu \phi - \f{ \alpha-2X\beta }{2\alpha(\alpha+2X^2\beta_X-X\alpha_X)} ( \alpha_X g_{\rho\sigma} + \beta_X \partial_\rho \phi \partial_\sigma \phi) T^{\rho\sigma}_m  \partial^\mu \phi   } \\\notag
&+ \f{ T^{\rho\sigma}_m}{2} \kk{ \f{2X^2\alpha_X\beta_\phi-\alpha_\phi(\alpha+2X^2\beta_X)}{\alpha(\alpha+2X^2\beta_X-X\alpha_X)}  g_{\rho\sigma} + \f{\partial_\rho \phi \partial_\sigma \phi }{\alpha+2X^2\beta_X} \mk{ -\beta_\phi + \f{X\beta_X[ 2X^2 \alpha_X \beta_\phi - \alpha_\phi (\alpha+2X^2\beta_X) ]}{\alpha(\alpha+2X^2\beta-X\alpha_X)} } }=0.
\end{align}
If $T^{\mu\nu}_m$ does not contain any second order derivatives of
$g_{\mu\nu}$, then we see that \eqref{eq1s} and \eqref{eq2s} both contain at most second order derivatives of $g_{\mu\nu}$.  Moreover, they both contain at most third order derivatives of $\phi$, which result from the second order derivatives of $\tilde g_{\mu\nu}$ appearing in $\mathcal E^{\mu\nu}_H$ and $\mathcal E_H^{(\phi)}$.  Note that, as a result of the kinetic
mixing between the scalar field and the metric in \eqref{eq2}, the
equation of motion \eqref{eq2s} contains mixing terms between
$\phi$ and the matter energy-momentum tensor and its derivatives.  The
rich structure of kinetic mixing found in general scalar-tensor theories
has many interesting consequences, as discussed in \cite{Bettoni:2015wta}.

Although the situation has been marginally improved in moving from
\eqref{eq1} and \eqref{eq2} to \eqref{eq1s} and \eqref{eq2s}, we still
have third order derivatives of $\phi$ appearing, which would normally
indicate the presence of dangerous additional degrees of freedom.  One
can superficially remove these third order derivatives by using
the trace of the gravitational equations \eqref{eq1s} -- sometimes referred to as a ``hidden constraint'' -- to find an expression for
$\dddot\phi$ in terms of lower order derivatives of $\phi$ and
$g_{\mu\nu}$, which can then be substituted into \eqref{eq1s}
\cite{Zumalacarregui:2013pma,Gleyzes:2014qga}.  However, as recently
highlighted in \cite{Deffayet:2015qwa}, this does not offer a rigorous
proof that the number of degrees of freedom is unaffected by the
disformal transformation, as the trace equation itself 
is a dynamical equation that must still be satisfied.

A more rigorous approach, which is in the same vein as that mentioned above and has been demonstrated in \cite{Deffayet:2015qwa}, is to find a suitable linear
combination of
the gravitational equations of motion \eqref{eq1s} that 
does not contain $\dddot\phi$ and
allows you to
express $\ddot\phi$ in terms of at most first order time derivatives of
$\phi$ and $g_{\mu\nu}$.  On taking the derivative of this relation we
are then able to find an expression for $\dddot\phi$ in terms of at most
second order time derivatives of $\phi$ and $g_{\mu\nu}$, which can be
substituted back into the original equations.  The difference here is
that the special combination of gravitational equations used does not
itself involve third order derivatives of $\phi$.  In
\cite{Deffayet:2015qwa}, Deffayet {\it et al.}\ showed that such a
combination does indeed exist for a very general class of models in the
context of covariantized Galileons.  
More specifically, they considered theories that in four spacetime dimensions are equivalent to \eqref{hornAct} but with the replacements $G_{4X}(\phi,X) \to F_4(\phi,X)$ and $G_{5X}(\phi,X) \to F_5(\phi,X)$, where $F_4$ and $F_5$ are arbitrary functions of $\phi$ and $X$.  When evaluated in the unitary gauge, this Lagrangian is equivalent to the Lagrangian of the theories beyond Horndeski's considered in \cite{Gleyzes:2014dya}, and as mentioned in the introduction, the structure of this Lagrangian is preserved under disformal transformations with $\alpha=\alpha(\phi)$ and $\beta=\beta(\phi,X)$.  
The models under consideration here, however, are related to Horndeski's theory by a more general disformal transformation with $\alpha=\alpha(\phi,X)$ and $\beta=\beta(\phi,X)$, meaning that they do not belong to the class analysed by Deffayet {\it et al.}  As such, the appropriate combination of gravitational equations with which we can express $\ddot \phi$ in terms of first order time derivatives of $\phi$ and $g_{\mu\nu}$ remains to be found.

Another rigorous approach to determine the number of degrees of freedom
of the theory is to perform a Hamiltonian analysis.  Such an analysis
has been performed for the newly-discovered theories beyond Horndeski's in 
Refs.~\cite{Gleyzes:2014qga,Lin:2014jga,Gao:2014fra}, but in these analyses
the unitary gauge was taken from the outset, which could affect the
outcome of the degrees-of-freedom counting \cite{Deffayet:2015qwa}.  A
Hamiltonian analysis that does not rely on fixing the gauge has been
performed for an example Lagrangian in \cite{Deffayet:2015qwa}, but its
generalisation is yet to be carried out.

In the context of theories that are disformally related to Horndeski's,
we are particularly interested in how the Hamiltonian analysis of a
theory is affected by a disformal transformation.  In the case of a
conformal transformation, where $\beta=0$ and $\alpha =
\alpha(\phi)$, it has been demonstrated in the context of $f(R)$ gravity
that the transformation simply gives rise to a canonical
transformation of variables, thus rendering the Hamiltonian analysis
unchanged \cite{Deruelle:2009pu}.  More general metric transformations -- which include disformal transformations of the form \eqref{disTran} --
were also considered recently in \cite{Domenech:2015tca}, where they
reached a similar conclusion; we will comment further on their results shortly.  Whilst a complete analysis of how the
Hamiltonian analysis of a theory is affected by disformal
transformations is beyond the scope of this paper, it is nevertheless
possible to see why we might expect the degrees-of-freedom counting to
remain unchanged.

The key distinguishing feature of 
general disformal transformations taking the form \eqref{disTran} 
as opposed to conformal transformations
with $\beta = 0$ and $\alpha= \alpha(\phi)$ is that they induce higher
order derivatives of some of the fields in the theory at the level of
the action.\footnote{To be more specific, we are interested in the case where both $\alpha$ and $\beta$ have an $X$-dependence.  As discussed previously, in the case where $\alpha = \alpha(\phi)$ and $\beta = \beta(\phi)$ Horndeski's theory is mapped onto itself, and we already know that Horndeski's theory propagates only three degrees of freedom.  Similarly, in the case where $\alpha = \alpha(\phi)$ and $\beta = \beta(\phi,X)$, Horndeski's theory will map onto a subclass of the theories introduced in \cite{Gleyzes:2014dya}, which were shown to be healthy without making use of the unitary gauge in \cite{Deffayet:2015qwa}.}  However, the fact that these higher order derivatives
always appear in a certain combination -- as we will see with the help
of a toy model below -- means that they are always associated with
additional primary constraints in the Hamiltonian analysis.  If these
constraints are first class, or second class and give rise to
secondary second class constraints, then they will remove the spurious
degrees of freedom associated with the higher derivatives appearing in
the action as a result of the transformation.  Indeed, this is the case
in the example model considered in Sec.\ III of \cite{Domenech:2015tca},
where a derivative of the lapse, $N$, induced by the disformal
transformation is always associated with a derivative of the spatial
metric, $\gamma_{ij}$, leading to a primary constraint involving the
corresponding canonical momenta $\pi_N$ and $\pi^{ij}$.  The
``exorcising of Ostrogradsky's ghost'' with constraints was also
discussed in \cite{Chen:2012au}.

In order to demonstrate the above idea, let us consider a simple toy
model consisting of two degrees of freedom $x(t)$ and $z(t)$, with
$\mathcal L_0 = \mathcal L_0(x,\dot x, z, \dot z)$.  Here we have in
mind that $x$ represents the metric degrees of freedom $\tilde
g_{\mu\nu}$ while $z$ the scalar field $\phi$.  Assuming this Lagrangian
to be regular means that we have two degrees of freedom and require four
initial conditions.  Analogous to a disformal transformation, we then
consider that $x = x(y,z,\dot z)$, where $y(t)$ represents the
disformally related metric $g_{\mu\nu}$, and the transformation depends
on the analogues of $\phi$ and $X$, $z$ and $\dot z$ respectively.  In
terms of this new set of variables we have $\mathcal L_0 = \mathcal
L_0(y,\dot y, z, \dot z, \ddot z)$, i.e.\ due to the appearance of $\dot
z$ in the transformation we have picked up a dependence of $\mathcal
L_0$ on $\ddot z$.  Following the constrained Ostrogradsky approach to
systems with higher order derivatives -- see e.g.\
\cite{Govaerts:1994zh,Querella:1998ke} -- we introduce an auxiliary
degree of freedom as $w =\dot z$, which we impose by adding a Lagrange
multiplier to the Lagrangian, i.e.\ we have
\begin{equation}
 \mathcal L = \mathcal L_0(y,\dot y, z, w,\dot w) + \lambda(w - \dot z).
\end{equation}
This Lagrangian depends on four degrees of freedom and contains up to
first order derivatives, so in general eight initial conditions are
required.  On calculating the canonical momenta we obtain
\begin{equation}
 p_y = \frac{\partial \mathcal L_0}{\partial\dot y}, \qquad p_z = -\lambda,\qquad p_w = \frac{\partial \mathcal L_0}{\partial \dot w} \qquad \mbox{and}\qquad p_\lambda = 0, 
\end{equation}
meaning that we have two
obvious primary constraints, namely
\begin{equation}
 \Phi_1 = p_z + \lambda \approx 0\qquad\mbox{and}\qquad \Phi_2 = p_\lambda \approx 0. \end{equation}
Note that these two constraints result from the introduction of the auxiliary degree of freedom, and in this sense do not correspond to the ``additional primary constraints'' referred to above.  In order to see how these additional primary constraints appear we note that
\begin{equation}
 p_y = \frac{\partial\mathcal L_0}{\partial \dot y} = \frac{\partial \mathcal L_0}{\partial \dot x}\frac{\partial \dot x}{\partial \dot y} = \frac{\partial\mathcal L_0}{\partial \dot x}\frac{\partial x}{\partial y}\qquad \mbox{and}\qquad  p_w = \frac{\partial\mathcal L_0}{\partial \dot w} = \frac{\partial \mathcal L_0}{\partial \ddot z}= \frac{\partial \mathcal L_0}{\partial \dot x}\frac{\partial \dot x}{\partial \ddot z} = \frac{\partial\mathcal L_0}{\partial \dot x}\frac{\partial x}{\partial \dot z},
\end{equation}
so that we find the additional primary constraint 
\begin{equation}
 \Phi_3 = p_y - F(y,z,w)p_w\approx 0,\qquad \mbox{where} \qquad F(y,z,w) \equiv \frac{\partial x}{\partial y}\frac{1}{\partial x/\partial \dot z}.
\end{equation}
On constructing the total Hamiltonian and requiring that the constraints $\Phi_i$ be preserved under time evolution we find an additional secondary constraint, which we label $\Phi_4$.  Requiring $\Phi_4$ to be preserved under time evolution does not give rise to any additional secondary constraints, and we find that all four constraints are second class.  As such, although we started with four degrees of freedom requiring eight
initial conditions, we found four second class constraints,
meaning that we only have to specify four initial conditions,
corresponding to only two degrees of freedom.  This is in agreement with
the fact that we had two degrees of freedom present in the original Lagrangian $\mathcal
L_0(x,\dot x, z, \dot z)$.

The key point in the above analysis was the appearance of the third
primary constraint $\Phi_3$.  As alluded to earlier, this arose due to
the fact that $\ddot z$ only appears in the Lagrangian in a specific
combination with $\dot y$, which is a consequence of the form of the
analogue of the disformal transformation $x = x(y,z,\dot z)$.

In fact, the toy model
considered above falls into the more general class of models and
transformations considered in \cite{Domenech:2015tca}.  To see this we introduce two auxiliary degrees of freedom, which allows us to re-write the Lagrangian in a form that is linear in velocities, i.e.\ we write
\begin{equation}\label{domrep}
 \mathcal L = \mathcal L_0(x,u,z,w) + \lambda_1(u-\dot x) + \lambda_2(w-\dot z).
\end{equation}
When written in terms of $w$, the transformation no longer depends on the velocity of any of the fields, i.e.\ we have $x = x(y,z,w)$.  This form of transformation and the form of action given in \eqref{domrep} are then of the form considered by Dom\`enech {\it et al.}\ in \cite{Domenech:2015tca}.  Assuming that the transformation $x = x(y,z,w)$ is invertible, they show that such a transformation is a canonical transformation, so that the set of constraints and constraint algebra are left unchanged, which in turn means that the number of degrees of freedom remains the same.  Note that the analysis of Dom\`enech {\it et al.}\ does not rely on the original Lagrangian being regular, and so offers a general proof that the number of degrees of freedom is unaffected on making a disformal transformation, so long as the transformation is invertible.  
We thus conclude that theories disformally related to Horndeski's theory via an invertible disformal transformation of the form \eqref{disTran} are healthy, in the sense that they propagate only three degrees of freedom.

\section{Conclusion}
\label{sec:con}

In this paper we have examined how the comoving curvature perturbation
transforms under the general disformal transformations of the metric
given in \eqref{disTran}.  We began by considering linear perturbations,
and showed that whilst the vector and tensor perturbation are invariant,
the comoving curvature perturbation is not identically invariant under a
general disformal transformation.  The difference between disformally
related curvature perturbations is given in \eqref{zetaDiff}, and is
written in terms of the gauge-invariant comoving density perturbation
$\Sp$ associated with a single canonical scalar field.  In the context
of Horndeski's theory we used the Hamiltonian and momentum constraints
to derive a Poisson equation for $\Sp$, and found that a sufficient
condition for the vanishing of $\Sp$ is that $\R$ is conserved, which is
the case on superhorizon scales so long as the so-called decaying mode
of $\R$ can be neglected.  As such, we concluded that the comoving
curvature perturbation is disformally invariant on superhorizon scales
for any theory that is disformally related to Horndeski's theory,
provided that the decaying mode of $\R$ can be neglected.  Using this
result, we saw that under the same mild assumption the comoving density
perturbation is suppressed on superhorizon scales for any theory that is
disformally related to Horndeski's theory, which we were also able to
derive explicitly from the transformation rule for $\Sp$ under disformal
transformations.  The relation between $\tSp$ and $\Sp$, given in
\eqref{tsp}, shows that $\tSp$ is equal to $\Sp$ up to a proportionality
coefficient that depends on background quantities.  The coefficient is
finite so long as the condition given in \eqref{discond} for the
invertibility of the disformal transformation is satisfied.  Based on
these findings, we conclude that in most cases of interest in the
context of inflation, we are free to work in any disformally related
frame when we wish to calculate the superhorizon curvature and tensor
perturbations that are required in making predictions for inflationary
observables.  This is an extension of the conformal invariance of the
curvature and tensor perturbations that is frequently exploited in the
context of simple scalar-tensor theories, where it is often much easier
to perform calculations in one frame than in another.  One can therefore
expect that the disformal invariance of perturbations will also prove to
be very useful in this wider class of theories.

Using the unitary gauge, we also considered the comoving curvature
perturbation at full nonlinear order, and found that it is invariant
under disformal transformations if we assume that an attractor regime
has been reached, where $\partial_\tau\phi = f(\phi)$.  In the context
of Horndeski's theory, it is known that on superhorizon scales the
nonlinear curvature perturbation is conserved in this regime so long as
the so-called decaying mode can be neglected, which implies that on
superhorizon scales the nonlinear comoving curvature perturbation is
both disformally invariant and conserved in any theory that is
disformally related to Horndeski's, provided $\partial_\tau\phi =
f(\phi)$ and the decaying mode can be neglected.  In addition, we found
that the vector and tensor perturbation are invariant under a general
disformal transformation at the nonlinear level.

Finally, we discussed the healthiness of theories disformally related to
Horndeski's in relation to the presence of Ostrogradsky instabilities.
Focusing on a toy model, we saw that the appearance of higher
derivatives in an action that results from making a field transformation
involving time derivatives of some of the fields is accompanied by the
appearance of second class primary and secondary constraints in the Hamiltonian
analysis.  These constraints therefore remove the spurious additional
degrees of freedom so that, as one would naively expect, the number of
degrees of freedom is unaffected by the field redefinition.  This is in agreement with the results of a recent analysis by Dom\`enech {\it et al.}\ in \cite{Domenech:2015tca}, where it was shown that the number of degrees of freedom is indeed unaffected by disformal transformations, so long as the transformations are invertible.

\vspace{0.5cm}

\noindent {\bf Note added:} Our results are in agreement with those
presented in Ref.~\cite{Watanabe:2015uqa}, which appeared on
arXiv at the same time as the present article and contains some material that overlaps with the 
nonlinear analysis we presented in \S \ref{sec:non}.

\acknowledgments

We would like to thank A.\ Naruko, M.\ Sasaki and Y.\ Watanabe for
useful discussions related to their work
\cite{Watanabe:2015uqa} prior to submission.  We would also like to
thank G.\ Dom\`enech, J.\ Gleyzes, M.\ Kar\v{c}iauskas, M.\ Minamitsuji, T.\ Suyama, 
D.\ Yamauchi and M.\ Zumalac\'{a}rregui for helpful discussions and comments.  
Finally, we would like to thank the anonymous referee, 
whose comments and questions have helped us to improve the manuscript.  
This work was partially supported by 
Japan Society for the Promotion of Science (JSPS)
Postdoctoral Fellowships for Research Abroad (H.M.) and JSPS
Grant-in-Aid for Scientific Research (B) No.~23340058 (J.W.).

\appendix

\section{Vanishing of $\Sp$ in Horndeski's theory}
\label{app:van}

In this appendix we derive the Poisson equation \eqref{genPoiss} in
Horndeski's theory, which is used in \S\ref{sec:lin} to argue that
under reasonable assumptions the comoving density perturbation $\Sp$,
defined in \eqref{comdp}, is suppressed on superhorizon scales.  This
result is a generalization of the result obtained in
\cite{Minamitsuji:2014waa} for a subclass of Horndeski's theory with
$G_4=1/2$, $G_5=0$.

Taking the action \eqref{hornAct}, assuming a flat FLRW background with line 
element of the form $ds^2 = -N^2(t)dt^2 + a^2(t)\delta_{ij}dx^idx^j$ and
varying the action with respect to $N(t)$ gives us one of the background
equations of motion \cite{DeFelice:2011hq}
\begin{equation}
 \mathcal E \equiv \sum_i\mathcal E_i =  0,
\end{equation}
where
\begin{align}
\mathcal E_2 &= 2XK_{X} - K,\\
\mathcal E_3 & = 6X\dot\phi HG_{3X}-2X G_{3\phi},\\
\mathcal E_4 &= -6H^2G_4+24H^2X(G_{4X}+XG_{4XX})-12HX\dot\phi G_{4\phi
 X} - 6H\dot\phi G_{4\phi},\\
\mathcal E_5 &= 2H^3X\dot\phi(5G_{5X}+2XG_{5XX})
 -6H^2X(3G_{5\phi}+2XG_{5\phi X}).
\end{align}
Whilst there are two additional background equations 
corresponding to a variation with respect to $a(t)$ and $\phi(t)$, 
we will not use them 
in the following analysis, so we omit them here.

Turning next to linear perturbations, 
we take a line element of the form given in \eqref{linPert}, 
but focus on the scalar perturbations.  In addition, we
fix the spatial gauge such that $E = 0$.  The equations of motion for
the perturbations are obtained by expanding the action \eqref{hornAct} to second
order and varying it with respect to each of the perturbation variables.
Here we will only need two of the equations -- the two constraint
equations obtained by varying the action with respect to $A$ and $B$ 
-- and they take the form \cite{DeFelice:2011hq}
\begin{gather}
\label{eomHorn1}
 -A_1\dot\psi + A_4 A + \frac{k^2}{a^2}\left(-A_3\psi - aBA_5\right) =
-A_2\dot{\delta\phi}-A_6\frac{k^2}{a^2}\delta\phi +\mu\delta\phi  ,  \\
\label{eomHorn2} 
-C_1\dot\psi + C_3A = -C_2\dot{\delta\phi}-C_4\delta\phi,
\end{gather}
where the coefficients are dependent on background quantities and are
given as 
\begin{align}\nonumber
 A_1 &= 6\Theta,\qquad 
 A_2 = -\frac{2(\Sigma + 3H\Theta)}{\dot\phi},\qquad 
 A_3 = 2\mathcal G_T,\qquad
 A_4 = 2\Sigma,\\\nonumber
 A_5 &= -2\Theta,\qquad 
 A_6 = \frac{2(\Theta - H\mathcal G_T)}{\dot\phi},\qquad 
 \mu= \mathcal E_{\phi},\\\label{C}
 C_1 &= 2\mathcal G_T,\qquad 
 C_2 = \frac{2(\Theta-H\mathcal G_T)}{\dot\phi},\qquad 
 C_3 = -2\Theta, \qquad 
 C_4 = \frac{1}{\dot\phi^2}\left[2(H\ddot\phi-\dot H\dot\phi)\mathcal G_T - 2\ddot\phi\Theta\right],
\end{align}
with $\Sigma$, $\Theta$ and $\mathcal G_T$ being defined as
\begin{equation}
 \Sigma = X\frac{\partial \mathcal E}{\partial X} +
 \frac{1}{2}H\frac{\partial \mathcal E}{\partial H},\qquad \Theta =
 -\frac{1}{6}\frac{\partial\mathcal E}{\partial H},\qquad \mathcal G_T =
 2\left[G_4 -2XG_{4X}-X\left(H\dot\phi
 G_{5X}-G_{5\phi} \right)\right].
\end{equation}

The easiest way to obtain \eqref{genPoiss} is to consider the unitary
gauge, where $\delta\phi = 0$.  
In this gauge we have $\psi = -\mathcal R_c$, $A = -\Sp/\dot\phi^2$ 
and $B = -(\Psi + \mathcal R_c)/(aH)$
and Eqs.~\eqref{eomHorn1} and \eqref{eomHorn2} thus reduce to
\begin{gather}
 A_1\dot{\mathcal R}_c - A_4\frac{\Sp}{\dot\phi^2} +
\frac{k^2}{a^2}\left(A_3\mathcal R_c + \frac{A_5}{H}(\Psi + \mathcal R_c)\right) = 0 , \\ 
\label{gInvMomCon}C_1\dot{\mathcal R}_c - C_3\frac{\Sp}{\dot\phi^2} = 0.
\end{gather}
Eliminating $\dot{\mathcal R}_c$ from these equations we arrive at
\eqref{genPoiss}.

One can, of course, derive \eqref{genPoiss} without having to fix the
gauge.  To see this explicitly, we eliminate $\dot\psi$ from
Eqs.~\eqref{eomHorn1} and \eqref{eomHorn2} to obtain
\begin{align} \label{Poieq}
 \left(A_2 - \frac{A_1C_2}{C_1}\right)\dot{\delta\phi} + \left(A_4
 -\frac{A_1C_3}{C_1}\right)A - \left(\frac{A_1C_4}{C_1} +
 \mu\right)\delta\phi = \frac{k^2}{a^2}\left(A_3\psi + aBA_5 - A_6\delta\phi\right).
\end{align}
Considering the left-hand side (l.h.s.) of \eqref{Poieq} first, it can be re-written as 
\begin{align}\label{lhs}
 {\rm l.h.s.} =
 \frac{1}{\dot\phi^2}\left(\frac{A_1C_3}{C_1}-A_4\right)\left[\dot\phi^2\left(\frac{A_2C_1
 -A_1C_2}{A_1C_3-A_4C_1}\right)\dot{\delta\phi} -\dot\phi^2 A -
 \dot\phi^2\left(\frac{A_1C_4 + \mu C_1}{A_1C_3-A_4C_1}\right)\delta\phi\right].
\end{align}
Using \eqref{C} one can then confirm that
\begin{align}
\frac{A_2C_1-A_1C_2}{A_1C_3-C_1A_4} = \frac{1}{\dot\phi}.
\end{align}
In evaluating the coefficient of the $\delta\phi$ term we note that as a
result of the background equation of motion $\mathcal E = 0$ we have
\begin{align}
 \frac{d\mathcal E}{dt} = \frac{\partial\mathcal
 E}{\partial\phi}\dot\phi + \frac{\partial\mathcal E}{\partial X}\dot X
 + \frac{\partial\mathcal E}{\partial H}\dot H = 0,
\end{align}
from which we are able to obtain an expression for
$\mu=\mathcal E_{\phi}$.
Using this result, and noting that $\dot X = 2X\ddot\phi/\dot\phi$, we find
\begin{align}
 \frac{A_1C_4 +\mu C_1}{A_1C_3-A_4C_1} = \frac{\ddot\phi}{\dot\phi^2}.
\end{align}
Altogether we thus obtain
\begin{align}
 {\rm l.h.s.} =
 \frac{1}{\dot\phi^2}\left(\frac{A_1C_3}{C_1}-A_4\right)\Sp.
\end{align}
Turning next to the right-hand side (r.h.s.) of \eqref{Poieq}, we are able to re-write it as
\begin{align}
 {\rm r.h.s.} = -\frac{k^2}{a^2}\left[A_3\mathcal R_c
 +\frac{A_5}{H}\left(\Psi + \mathcal R_c\right) +
 \left(\frac{A_3H}{\dot\phi}+A_6 + \frac{A_5}{\dot\phi}\right)\delta\phi\right].
\end{align}
Using \eqref{C} one
can then show that $A_3H + \dot\phi A_6 + A_5 = 0$.  As such, we have
recovered \eqref{genPoiss} without fixing the time slicing.

\section{Equivalence of equations of motion in disformally related frames}
\label{app:equiv}

In this appendix, following the analyses of \cite{Deruelle:2014zza,Arroja:2015wpa}, we show that the gravitational equations given in \eqref{eq1} are equivalent to the gravitational equations of motion determined by varying \eqref{genact} with respect to $\tilde g_{\mu\nu}$, so long as the disformal transformation relating $\tilde g_{\mu\nu}$ and $g_{\mu\nu}$ is invertible,
namely the invertibility condition \eqref{discond} is satisfied.

We will use the following relation 
between $T^{\mu\nu}_m$ and $T^{\mu\nu}_H$, i.e.\ the inverse relation of \eqref{Tmunurel}, which is given as 
\begin{equation}
 T^{\mu\nu}_m = \alpha\sqrt{\alpha(\alpha - 2X\beta )}\left\{\alpha T^{\mu\nu}_H + \frac{\partial^\mu\phi\partial^\nu\phi}{2}\left(\alpha_Xg_{\alpha\beta} + \beta_X\partial_\alpha\phi\partial_\beta\phi\right)T^{\alpha\beta}_H\right\}.
\end{equation}
Here, the coefficient in front of the large bracket on the right hand side does not vanish as \eqref{discond} is satisfied.
Using the above relation it is possible to rewrite \eqref{eq1} as 
\begin{equation}\label{rweq1}
 \alpha\left(\mathcal E_H^{\mu\nu}+\frac{1}{2}T^{\mu\nu}_H\right) +\frac{1}{2}\left(\mathcal E_H^{\rho\sigma} + \frac{1}{2}T^{\rho\sigma}_H\right)\left(\alpha_X g_{\rho\sigma}+\beta_X\partial_\rho\phi\partial_\sigma\phi\right)\partial^\mu\phi\partial^\nu\phi = 0.
\end{equation}
Contracting this equation once with $g_{\mu\nu}$ and once with $\partial_\mu\phi\partial_\nu\phi$ we obtain the two equations
\begin{align}
 &(\alpha -X\alpha_X )P -X\beta_X Q = 0,\\
 &2X^2\alpha_X P + (\alpha+2X^2\beta_X )Q = 0,
\end{align}
where
\begin{equation}
 P = \left(\mathcal E_H^{\mu\nu}+\frac{1}{2}T^{\mu\nu}_H\right)g_{\mu\nu}\qquad\mbox{and}\qquad 
 Q = \left(\mathcal E_H^{\mu\nu}+\frac{1}{2}T^{\mu\nu}_H\right)\partial_\mu\phi\partial_\nu\phi.
\end{equation}
The solution to these two equations is $P=0$ and $Q = 0$, so long as the matrix $M$, given as 
\begin{equation}
 M = \begin{pmatrix}
  \alpha-X\alpha_X  & -X\beta_X \\ 
  2X^2\alpha_X & \alpha+2X^2\beta_X
  \end{pmatrix}
\end{equation}
is invertible.  Substituting $P=Q=0$ into \eqref{rweq1}, and assuming $\alpha\ne 0$, we thus recover
\begin{equation}
 \mathcal E_H^{\mu\nu}+\frac{1}{2}T^{\mu\nu}_H=0,
\end{equation} 
i.e.\ we have recovered the gravitational equations obtained on minimising \eqref{genact} with respect to $\tilde g_{\mu\nu}$.  
The case when $M$ is not invertible corresponds to 
\begin{equation}
 \alpha(\alpha - X \alpha_X  + 2X^2\beta_X) = 0, 
\end{equation}
meaning that we require $\alpha \ne 0$ and $\alpha -X \alpha_X + 2X^2\beta_X \ne 0$,
which is the case so long as the invertibility condition \eqref{discond} is satisfied.  

\bibliography{refs}
\end{document}